%% file: manuscript.tex
\documentclass[10pt,conference]{IEEEtran}
\IEEEoverridecommandlockouts
\usepackage{cite}

\input{packages}

\begin{document}

\title{
  Quantum-safe Edge Applications: How to Secure Computation in Distributed Computing Systems
}

\author{
\IEEEauthorblockN{Claudio Cicconetti}
\IEEEauthorblockA{IIT-CNR \\
Pisa, Italy}
\and
\IEEEauthorblockN{Dario Sabella}
\IEEEauthorblockA{Intel Corporation\\
Turin, Italy}
\and
\IEEEauthorblockN{Pietro Noviello}
\IEEEauthorblockA{Exprivia S.p.A. \\
Molfetta, Italy}
\and
\IEEEauthorblockN{Gennaro Davide Paduanelli}
\IEEEauthorblockA{Exprivia S.p.A. \\
Molfetta, Italy}
}

\IEEEtitleabstractindextext{%
\begin{abstract}
The advent of distributed computing systems will offer great flexibility for application workloads, while also imposing more attention to security, where the future advent and adoption of quantum technology can introduce new security threats. For this reason, the Multi-access Edge Computing (MEC) working group at ETSI has recently started delving into security aspects, especially motivated by the upcoming reality of the MEC federation, which involves services made of application instances belonging to different systems (thus, different trust domains). On the other side, Quantum Key Distribution (QKD) can help strengthen the level of security by enabling the exchange of secure keys through an unconditionally secure protocol, e.g., to secure communication between REST clients and servers in distributed computing systems at the edge. In this paper, we propose a technical solution to achieve this goal, building on standard specifications, namely ETSI MEC and ETSI QKD, and discussing the gaps and limitations of current technology, which hamper full-fledged in-field deployment and mass adoption. Furthermore, we provide our look-ahead view on the future of secure distributed computing through the enticing option of federating edge computing domains.
\end{abstract}

}

\maketitle

\begin{tikzpicture}[remember picture,overlay]
\node[anchor=south,yshift=10pt] at (current page.south) {\fbox{\parbox{\dimexpr\textwidth-\fboxsep-\fboxrule\relax}{
  \footnotesize{
     \copyright 2024 IEEE.  Personal use of this material is permitted.  Permission from IEEE must be obtained for all other uses, in any current or future media, including reprinting/republishing this material for advertising or promotional purposes, creating new collective works, for resale or redistribution to servers or lists, or reuse of any copyrighted component of this work in other works.
  }
}}};
\end{tikzpicture}%

\IEEEdisplaynontitleabstractindextext

\IEEEpeerreviewmaketitle


\section{Introduction}\label{sec:introduction}

\begin{figure}[tb]
    \centering
    \includegraphics{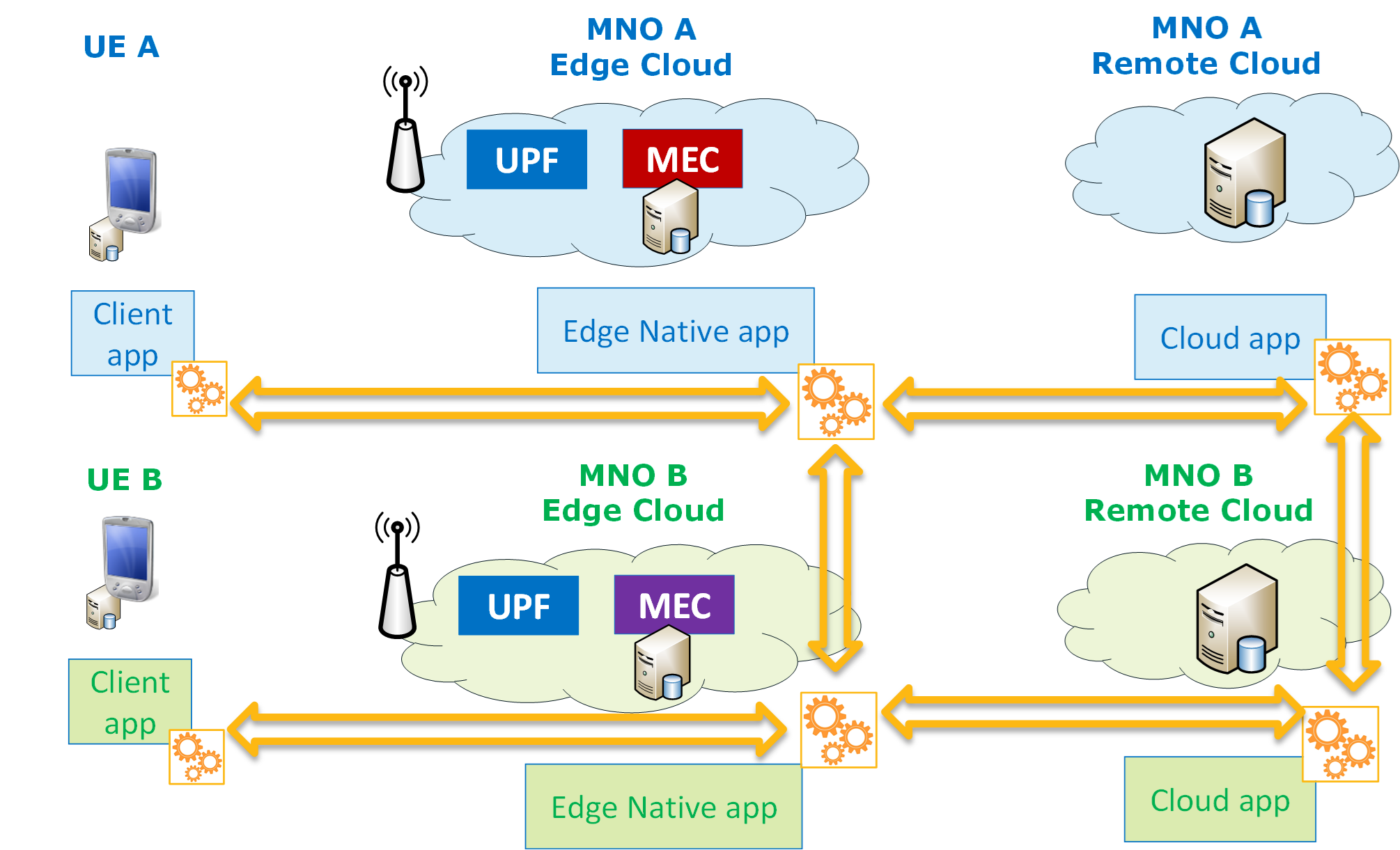}
    \caption{Distributed computing scenarios}
    \label{fig:distr-comp-scenarios}
    \vspace{-1em}
\end{figure}

Distributed computing scenarios offer a variety of computing partitioning, with the advantage of potentially optimizing the distribution of the workload to minimize costs and maximize performance~\cite{chen_towards_2023}.
In this context, standardization of the edge infrastructure ensures interoperability, where \emph{federation}, as defined by ETSI MEC~\cite{ETSIMEC003}, can be a convenient framework to enable this kind of heterogeneous deployments, as shown in \Cref{fig:distr-comp-scenarios}.
Securing communications in those complex scenarios is however critical, and quantum cryptography and \ac{QKD} can help in strengthening the level of security of these environments.

QKD is the most mature among the Quantum Information Technologies, which are all advancing at a fast pace to revolutionize both computing, through a speed-up of computation to solve problems that are intractable with classical (i.e., non-quantum) computers~\cite{quantum_technology_and_application_consortium__qutac_industry_2021}, and communication, with the deployment of the Quantum Internet~\cite{gyongyosi_advances_2022}.
However, despite some ongoing efforts to build quantum networks at a geographical scale (e.g., the EuroQCI initiative in EU\footnote{\url{https://digital-strategy.ec.europa.eu/en/policies/european-quantum-communication-infrastructure-euroqci}}), the goal of building a global infrastructure in the next years is too ambitious, mainly because of two reasons.
First, quantum communications rely on the encoding of quantum states in single photons, which makes the transmission very fragile, especially at long distances.
Second, quantum bits cannot be copied/amplified, due to the ``no-cloning'' theorem of quantum mechanics, which makes the manufacturing of so-called quantum repeaters overly complex~\cite{rakonjac_storage_2022}.
For this reason, it is generally assumed that, at least in the near future, quantum networks will be limited to Quantum Metropolitan Area Networks (QMANs), which makes them an excellent candidate for being used in tandem with edge computing, possibly with the help of \acp{TN}~\cite{cao_evolution_2022}.
A \ac{TN} is a device that interconnects two QKD devices and exchanges secret keys with both of them to create a logical tunnel for the secure end-to-end exchange of keys, which however is only as secure as the \ac{TN} itself.

The remainder of this paper is structured as follows.
In \Cref{sec:use-case} we illustrate two example use cases that motivate the need for secure edge computing.
The interworking of ETSI MEC and QKD to achieve this goal is described in detail in \Cref{sec:integration}, which is followed by a high-level look ahead on edge computing federation in \Cref{sec:look-ahead}.
\Cref{sec:conclusions} concludes the paper.

\section{Motivating Use Cases}\label{sec:use-case}


\subsection{Cybersecurity in Healthcare}\label{sec:use:health}

Professional healthcare is undergoing major advances thanks to the broader adoption of \ac{ICT}, especially through the use of \ac{AI} algorithms to help doctors in the diagnosis phase, for emergency and evaluation of triage, and for continuously monitoring patients.
Hospitals, sanitaria, and long-term care facilities are increasing the adoption of body/room sensors for the collection of vital parameters, which can then be used as input for inference models based on \ac{ML}-trained \acp{NN}.
The latter can be very computation-intensive and, thus, require high-performance computing infrastructures that are not often available on-site.
Such data are highly confidential and protected by many national regulations (e.g., General Data Protection Regulation (GDPR) in the EU member states).
This makes a strong case for using QKD and edge computing~\cite{ur_rasool_quantum_2023}: the security brought by the former is reinforced by the latter, which keeps the remote interactions as close as possible to where the data are generated/consumed.
In this use case, the administrative boundaries (or security perimeters), both hosting QKD devices interconnected through a quantum network, would be the healthcare facilities on the user side, and the edge computing infrastructure on the telco operator side.

\subsection{Cybersecurity in Automotive}\label{sec:use:automotive}

\begin{figure}[tb]
    \centering
    \includegraphics{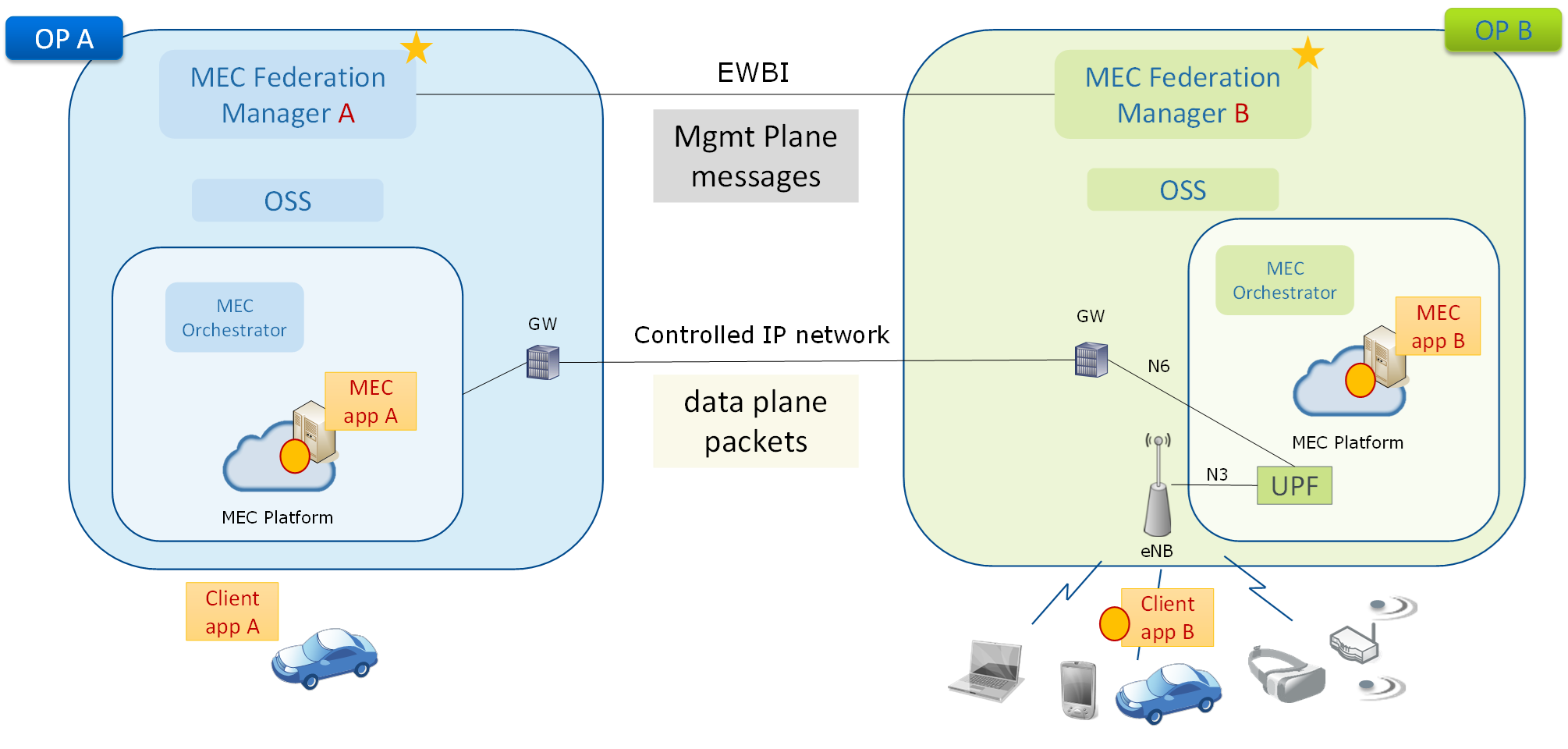}
    \caption{Exemplary deployment of a MEC federation}
    \label{fig:federation}
    \vspace{-1em}
\end{figure}

Automotive and smart city scenarios in real life are characterized by complex environments with multiple mobile network operators, vendors, and integrators.
From an application perspective, these environments are populated by several types of workloads, i.e. client applications, MEC applications and backend applications~\cite{5gaa}.
In this context, an automotive scenario can be realized with the multi-stakeholder architecture depicted in \Cref{fig:federation}: the figure also exemplifies one of the key use cases for the MEC Federation, where multiple OP instances (as defined by GSMA OPG~\cite{gsma_opg}) are connected via EWBI (East-West Bound Interface) \acp{API}, for control plane traffic~\cite{ETSIMEC040}, and via a controlled IP network, for the data plane traffic.
The improved security with QKD networks/technologies would increase acceptance, a critical barrier to market penetration for this use case.
Due to the mobile nature of vehicular users, QKD devices cannot be co-located with individual terminals; instead, the QKD devices, interconnected to the quantum network via a fiber optic infrastructure, can be deployed in the security perimeter of the vehicular network, e.g., on Road Side Units (RSUs).

\section{Interworking of MEC and QKD}\label{sec:integration}

\begin{figure}[tb]
    \centering
    \includegraphics[scale=1]{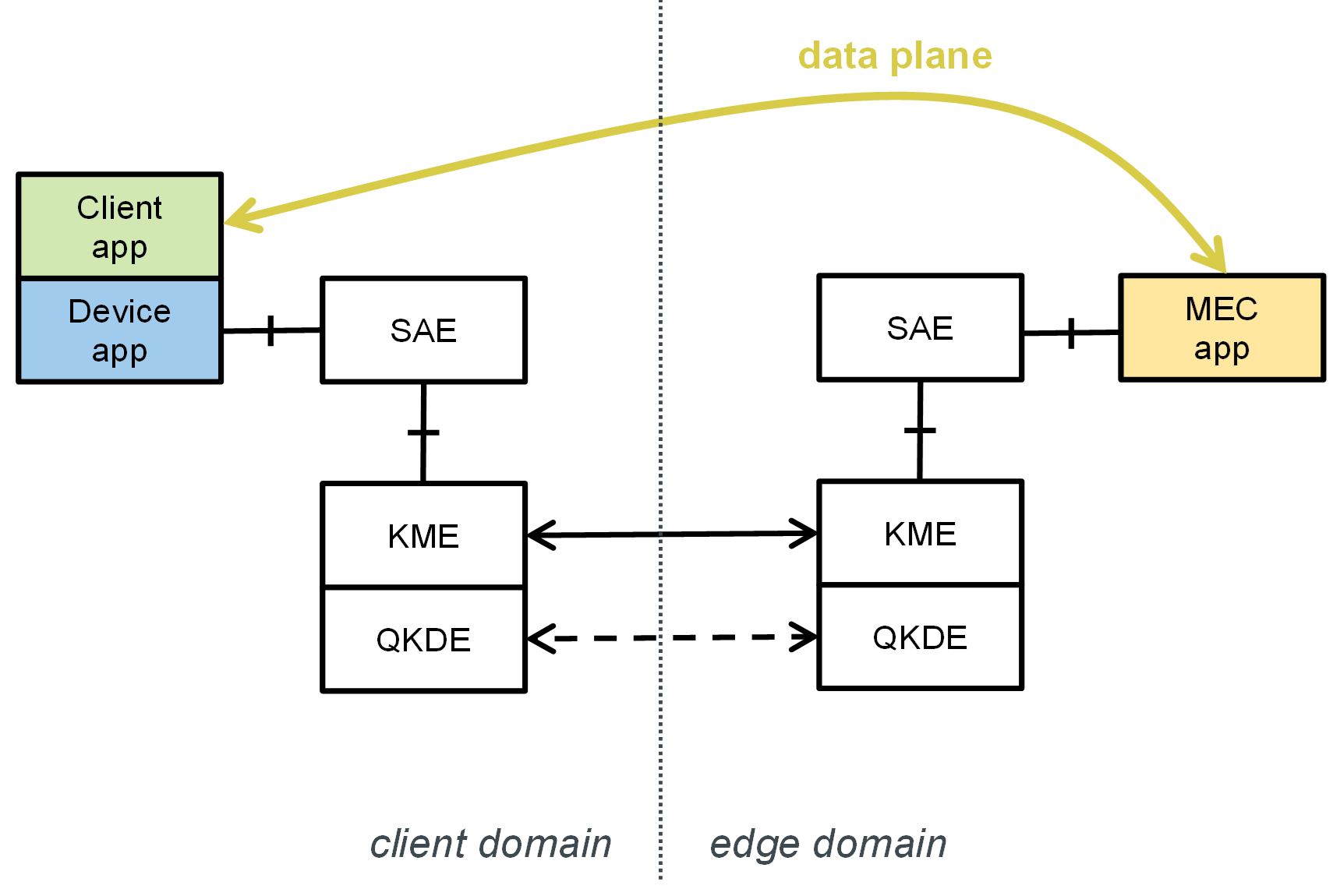}
    \caption{High-level interworking between ETSI MEC and QKD.}
    \label{fig:inter-high-level}
    \vspace{-1em}
\end{figure}

At a high level, the proposed solution to integrate QKD with an edge domain, where resources are deployed and managed through ETSI MEC, is illustrated in \Cref{fig:inter-high-level}.
The diagram shows a client and an edge domain, which are interconnected by both classical (5G and beyond) and quantum networks.
By design, the application on the client terminal (\emph{client app}) is transparent to both ETSI MEC and ETSI QKD procedures: this enables a gradual transition of customer applications towards the final goal and reduces the business risks related to the possible emergence of future alternative standards/frameworks.
Using ETSI MEC terminology, the client app interacts with a \emph{device app}, which is aware of the ETSI MEC procedures and, in particular, supports the client-side protocol of the \texttt{Mx2} interface defined in ETSI MEC 016~\cite{ETSIMEC016}.
The counterpart of the client app in the edge domain is called \emph{MEC app} and, in general, it is provided by the service provider to be run in the edge computing infrastructure as a partial or full substitute for cloud back-end services.
The device app exploits two capabilities offered by the ETSI MEC, both available through a component called \ac{LCMP} (not shown in this diagram): i) discovery, i.e., the device app may retrieve the list of MEC apps available on the edge domain, and ii) context creation, i.e., the device app can create an application context which, until its termination, will ensure that resources in the edge domain will be reserved and configured for the specific MEC app to which the context is attached.

Moving to the QKD segment, there are three main entities.
First, the \emph{\ac{QKDE}} is the QKD device, that is the termination of the underlying QKD network.
Through the continuous generation of locally entangled photons (on one side) and the measurement (on the other side), a pair of \acp{QKDE} can establish a stream of secret material that cannot be spoofed.
The \emph{\ac{KME}} is the component that combines such secret material in blocks and keys to make them available to the upper layers, also through a chain of intermediate \acp{TN} to extend beyond a local point-to-point connection between \acp{QKDE}~\cite{tysowski_engineering_2017}.
According to the ETSI QKD specifications, the \acp{KME} should cooperate to enable the discovery of available \acp{QKDE}, agree on the protocol for combining keys in a multi-hop fashion, and perform other management functions, but the specific procedures and interfaces are left unspecified.
On the other hand, the application API is defined in the ETSI QKD 014~\cite{ETSIQKD014}, which defines a REST protocol to be used by the applications to retrieve security key from a \emph{\ac{SAE}}, which in turn interacts with a local \ac{KME}.
The \ac{SAE}, \ac{KME}, and \ac{QKDE} should all reside within the same security perimeter to avoid the secret keys being compromised.

Once the ETSI MEC application context has been created, through the \texttt{Mx2} interface, the edge computing resources have been provisioned, and a security context has been created using the ETSI QKD 014 interface, the client and MEC apps can interact with one another by exchanging, through the classical network, encrypted messages that they only can decrypt with QKD-secure keys, as shown in the ``data plane'' bidirectional arrow in \Cref{fig:inter-high-level}.
In the remainder of this section, we provide additional details following a prototype implementation~\cite{cicconetti_prototype_2023}.

In the following, we assume that the client and MEC app interact through HTTP request/response messages.
This pattern is typical, for instance, of the \ac{FaaS} abstraction, often used in combination with serverless computing, which is a mature and very popular cloud-native technology~\cite{Khandelwal2021}.
With \ac{FaaS}, the service is broken down into elementary pieces that are invoked as functions in an edge/cloud serverless platform, either directly by the client or from other functions in a chain (or \ac{DAG}) of consecutive invocations, until the final result is returned to the client.
In this paper, we are interested only in encrypting the first and last step, i.e., the invocation of the client that triggers the execution of the function(s) and the last return value from the platform, since both may contain customer confidential data.
The intermediate invocations, if any, all happen within the edge domain, which is assumed to be within the same security perimeter, and therefore QKD-secure encryption is not necessary.
Furthermore, we are not concerned with the required exchanges within the QKD network to build secure keys across a chain of \acp{TN}, which do not affect the focus of our contribution and is a topic under study within the scientific literature (e.g., \cite{kong_challenges_2023}) and standardization bodies such as the ITU-T Focus Group on Quantum Information Technology for Networks (FG-QIT4N).
Therefore, without loss of generality, below we represent the QKD network as a point-to-point link.

The ETSI MEC components of the architecture are available as open-source code, programmed in Rust, under a permissive MIT license on GitHub\footnote{\url{https://github.com/ccicconetti/etsi-mec-qkd}}.

\subsection{Detailed Software Architecture}

\begin{figure*}[tb]
    \centering
    \includegraphics[width=\textwidth]{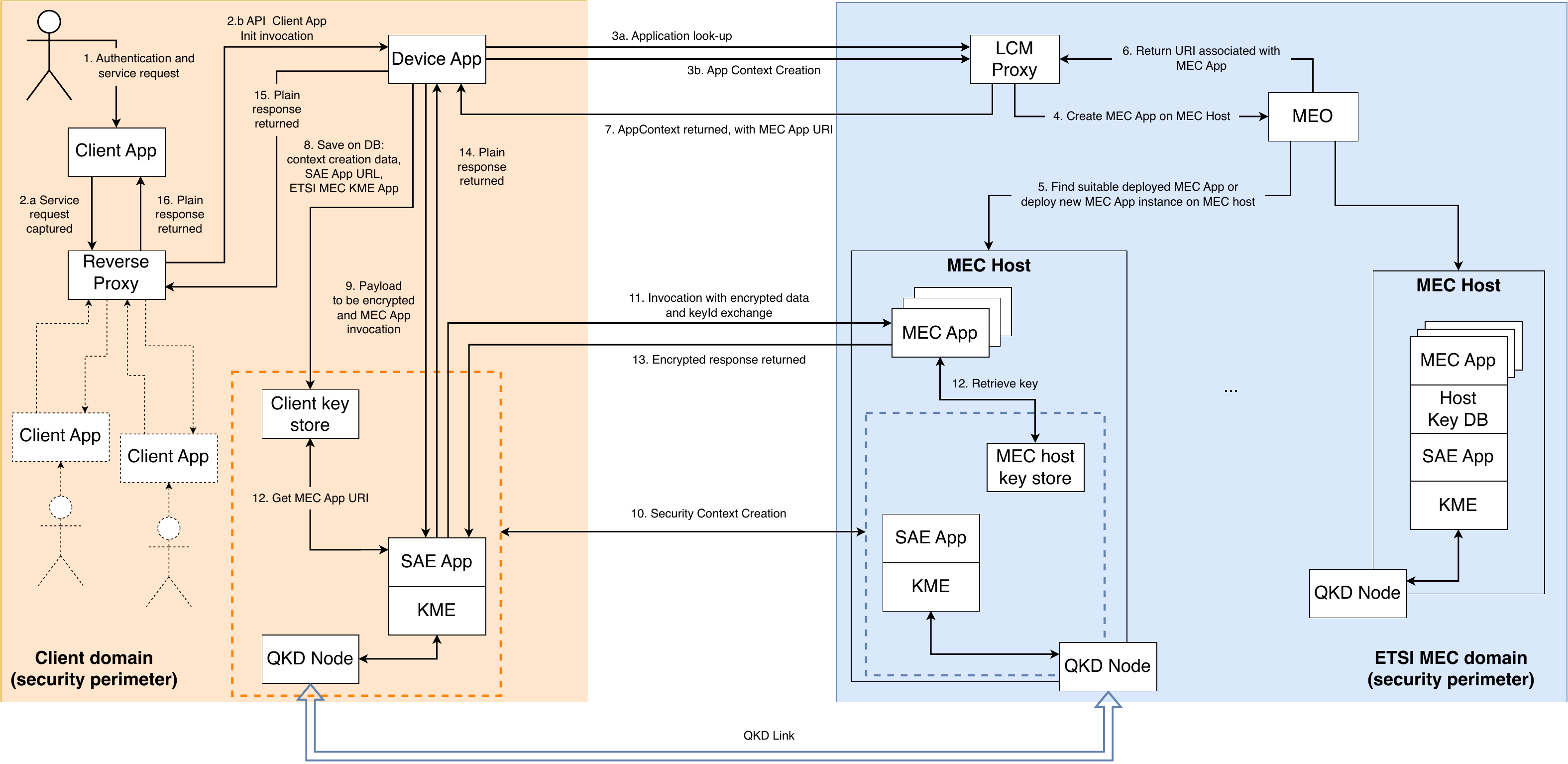}
    \caption{Detailed software architecture of an ETSI MEC computing infrastructure providing clients with services, also including encryption of private data with keys made available by a QKD network through an ETSI QKD 014 API. The security context creation procedure (step \textbf{10}) is shown separately in \Cref{fig:sec-context-creation}.}
    \label{fig:arch}
    \vspace{-1em}
\end{figure*}

A detailed software architecture of the proposed solution is illustrated in \Cref{fig:arch}, which shows additional components than those in the high-level view in \Cref{fig:inter-high-level}.
Specifically, in the client domain, in addition to the client app and device app, we find a \emph{reverse proxy}, which captures the HTTP requests/responses between the client app and the MEC app, to keep the former transparent to ETSI MEC/QKD procedures, and, the \emph{client key store}, which is an in-memory database that keeps data required for the encryption/decryption and context management.
Both the reverse proxy and the client key store can serve multiple clients within the same client domain, which may cover, for instance, a factory or a hospital served through a 5G private network and with a QKD node.
In the ETSI MEC domain, \Cref{fig:arch} shows the \ac{LCMP}, two \emph{MEC hosts}, which are the physical/virtual edge computing nodes dedicated by the telco operator to hosting MEC applications, the \emph{MEC host key store}, which is an in-memory database storing the keys used by the MEC app for encryption/decryption, and the \emph{\ac{MEO}}, which is the entity within an ETSI MEC domain in charge of managing the deployment and run-time optimization of the MEC apps.

In step \textbf{1} the user authenticates on their terminal via the local mobile app, which triggers the request for a service creation, issued by a REST command with HTTP and captured by the reverse proxy (\textbf{2a}).
The latter then directs an app initialization to the device app (\textbf{2b}), which is the entry point towards the ETSI MEC, also shared with other clients in the same domain.
The device app first performs an application look-up via the discovery features of the remote \ac{LCMP} (\textbf{3a}) to ensure that the type of MEC app requested is offered by a service provider hosted by the ETSI MEC domain, with matching version and capabilities as requested by the device app.
If this is the case, then the device app proceeds to request the creation of an app context to the \ac{LCMP} (\textbf{3b}).
The exchanges between the device app and the \ac{LCMP} are not subject to high-security constraints since no confidential data are exchanged, but traditional security means, such as \ac{TLS} can optionally be used.

The \ac{LCMP} then informs the \ac{MEO} of the request to create a new app context (\textbf{4}), which triggers a resource orchestration decision at the \ac{MEO}.
It looks for a suitable MEC app already deployed that can serve the new app context and, if not found, deploys a new MEC app on one of the MEC hosts with suitable characteristics, depending on the available resources and their locations, and the current status of utilization as obtained through the observability framework within the ETSI MEC domain.
In the literature, the problem of optimal allocation of resources in edge computing has been broadly investigated, typically to maximize resource utilization or minimize energy or latency (e.g., \cite{younis_energy-latency_2024,dai_offloading_2023}).
Once the MEC app is deployed, the \ac{MEO} can return the end-point to be used by the client app, through a \ac{URI}, to the \ac{LCMP} (\textbf{6}), which embeds it into the app context creation response returned to the device app (\textbf{7}).

At this point, the device app knows the end-point of the MEC app, which is cached in the client key store (\textbf{8}) and passes the original request from the client, with confidential data, to the SAE app.
The latter establishes a security context creation with the corresponding SAE app in the MEC host where the MEC app is deployed.
This procedure is illustrated separately later and it involves the QKD network underneath.
The security context creation concludes with the SAE app having a secret symmetric key with which it can encrypt the confidential client data before invoking the REST command on the MEC app listening at the given URI (\textbf{11}).
For instance, with serverless computing, such a command is a function invocation and the client data are passed as function arguments.
Upon receiving the invocation, the MEC app can retrieve its symmetric key from the MEC host key store (\textbf{12}), which plays the same role as that used by the SAE app for encrypting the client data, to decrypt and then consume the data.
It is worth noting that retrieving the key requires the identifier of the key (keyId), which was obtained during the security context creation procedure and passed by the SAE app as part of the invocation.
Once processing is complete, which may involve also the interaction with other MEC apps, not shown in the diagram, the final result is encrypted by the entry-point MEC app with the same key used for decrypting and returned to the SAE app (\textbf{13}).
The latter can decrypt the response and forward it to the device app (\textbf{14}), which in turn passes it to the reverse proxy (\textbf{15}) and, eventually, reaches the client app (\textbf{16}).

\subsection{Security context creation}

\begin{figure*}[tb]
    \centering
    \includegraphics[width=\textwidth]{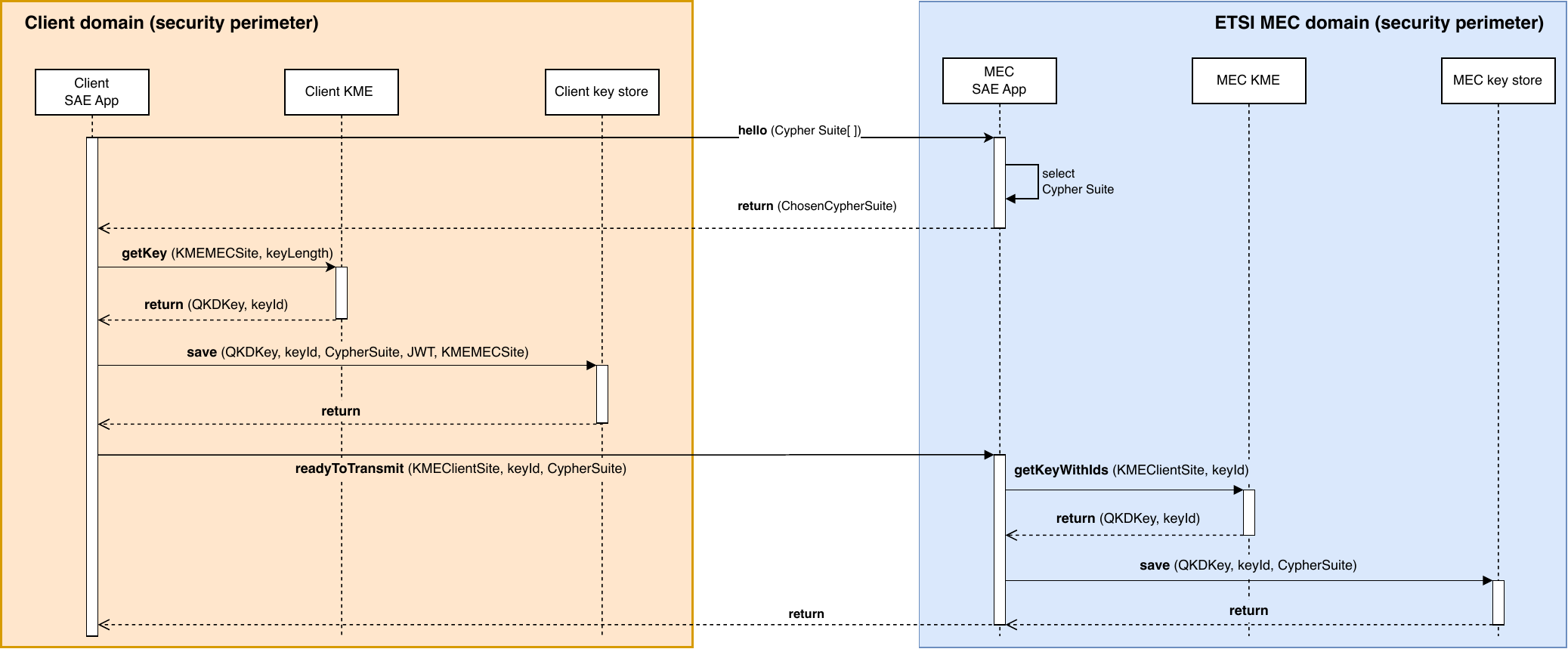}
    \caption{Security context creation procedure within the proposed software architecture (step \textbf{10} in \Cref{fig:arch}).}
    \label{fig:sec-context-creation}
    \vspace{-1em}
\end{figure*}

We now focus on the security context creation (step \textbf{10} in \Cref{fig:arch}), which is illustrated through the sequence diagram in \Cref{fig:sec-context-creation}.
In both domains, the three actors are the SAE App, the KME, and a key store.
The latter is used to cache keys and reuse them for multiple consecutive exchanges, as the KME, according to the ETSI QKD specifications, is not allowed to store secret material once it has been consumed by the SAE app.
Reusing the same key introduces a security threat since, theoretically, this makes it possible for an attacker to break the secret, but it allows more efficient use of the QKD resources, which are rather scarce with current technology (in the order of kbit/s~\cite{bacco_field_2019}).
The trade-off between security and efficiency in the selection of the key refresh rate depends on the specific characteristics and security targets of the application.

The procedure starts with the client advertising the set of known protocols, called CypherSuites in the diagram, to encrypt data to the MEC app, which selects a match (if any) and returns it.
This step determines, along with the algorithm, the length of the key to be used (keyLength).
The next step is done by the client, which uses the ETSI QKD 014 API to retrieve a key from its KME, also specifying an identifier of its peer, i.e., the MEC app.
Depending on the QKD network topology, the KME may have to negotiate with multiple KMEs for an end-to-end key to be created, but as already mentioned this step is not discussed in this paper, since it is transparent from the point of view of the SAE apps.
The KME then returns the client app a key with the given length and an identifier (keyId), and they are both saved in the client key store for later use.

The client app then sends to the MEC app the identifier of the client app, the keyId, and the CypherSuite.
With the identifier and the keyId, the MEC app can query its MEC KME via the ETSI QKD 014 API and retrieve the associated key, which is then cached in the key store of the MEC host for later use, similarly to how was done by the client.

\subsection{Discussion}

The use of QKD is subject to high deployment cost, in terms of both fiber optic cores that must be dedicated to QKD links and new equipment to be installed.
Therefore, it is tempting to evaluate the use of \ac{PQC} as a tout court alternative, as the latter merely requires software upgrades.
On the one hand, we can foresee that technology development will improve efficiency and reduce manufacturing costs of quantum devices, thus making QKD more cost-effective.
On the other hand, \ac{PQC}, while being a matter subject to intense research, will always be dependent on the computational power ascribed to the adversary, whereas QKD can be mathematically proven to be secure, independent of the resources of the adversary, being an Information Theory Secure (ITS) primitive~\cite{scarani_security_2009}.

Another deployment issue is that end-to-end security requires a QKD device to be installed in the security perimeter of the customer, which creates a new form of \emph{last mile} problem: if the customer cannot connect directly to the fiber-optic quantum network, then it is not possible to extend coverage through a copper-based or wireless 5G-like access network since they do not allow, at least today, the transmission of single photons encoded with a quantum state.
This is particularly relevant to scenarios where user terminals are mobile, such as in an automotive use case (\Cref{sec:use:automotive}).
One possibility under study is the use of free space through satellites, which is promising with both geo-stationary~\cite{yin_satellite-based_2017} and low-orbit~\cite{gundogan_proposal_2021} scenarios, though the related technology is lagging behind terrestrial fiber optic based quantum communications~\cite{gatto_integration_2022}.

Finally, in the solution illustrated above, the QKD network and the edge computing domain take independent resource management decisions: the former on the path to follow between the client and edge domains, the latter on which MEC host to use to deploy the MEC app.
This works well for demonstration purposes or with low adoption of the technology, but, in a fully grown scenario, it may lead to inefficient utilization of resources.
This can be overcome by a run-time management interaction between the QKD and edge computing operators, which is a topic understudied (despite some initial attempts, such as~\cite{cicconetti_qkdedge_2023}) and, to the best of our knowledge, not yet covered by standardization bodies.

\section{Looking ahead}\label{sec:look-ahead}

\begin{figure*}[tb]
    \centering
    \includegraphics[scale=1.75]{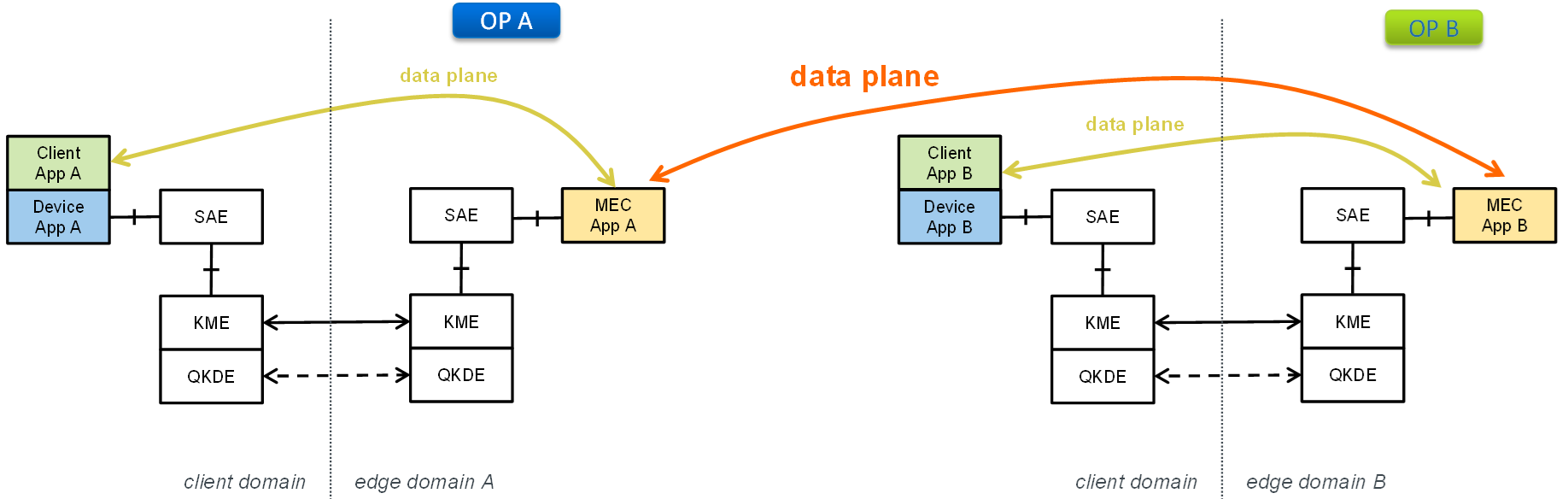}
    \caption{QKD-secure communication in MEC federation.}
    \label{fig:qkd-in-mec-fed}
    \vspace{-1em}
\end{figure*}

The federation of multiple infrastructures is a thriving topic in the edge-cloud ecosystem,
which multiplies the importance of security.
\Cref{fig:qkd-in-mec-fed} shows a possible communication between application instances with MEC federation, with proper security enabled by QKD in both MEC systems and the respective client domains.
However, even if the high-level concept is quite straightforward, proper standardization work is needed to allow actual implementations to fully exploit the benefits of QKD in MEC deployments.
In fact, according to ETSI GS QKD 014, the communication between applications in different security perimeters is assumed to be insecure, as performed via so-called \emph{public channels}.
This is consistent with the assumption in MEC systems that data plane connectivity between MEC apps is generically insecure.
Therefore, QKD-enabled secure communication between MEC apps with federation would be utterly needed.

As a starting point, ETSI QKD has already defined suitable QKD Key Servers\cite{ETSIQKD004}, which, at least in principle, can be conveniently deployed, e.g., co-located with the MEC Federator (MEF) or MEO, or stand-alone.
But many gaps need to be filled to achieve interoperability along the following directions, from the point of view of ETSI MEC:

\begin{enumerate}[parsep=0em,leftmargin=*,label=\arabic*.]
\item definition of signaling to allow protocol support for QKD-based authentication in a MEC federation $\rightarrow$ impact on authentication mechanisms defined in \textbf{ETSI MEC 009};
\item extension of the application profile to support exchanges between \acp{SAE} related to multiple MEC app instances $\rightarrow$ relevant for future releases of \textbf{ETSI MEC 011}, for example by properly enhancing the AppInfo data type by including an indication of the availability of a QKD Key Server in the system;
\item support in the MEC federators and the related flows $\rightarrow$ enhanced Federation Enablement APIs in \textbf{ETSI MEC 040}.
\end{enumerate}%

It is even possible to conceive ETSI MEC service APIs specifically to support secure communication between MEC applications in a federation, or adopting reference software implementations from relevant open-source communities.  

\section{Conclusions}\label{sec:conclusions}

In this paper, we have discussed a scenario for enabling QKD-secure data exchange in cloud-native applications deployed at the edge.
As a reference, we have considered a computing infrastructure based on the ETSI MEC architecture, and we have assumed that the quantum devices offer ETSI QKD 014 APIs.
Furthermore, we have discussed some key threats and opportunities for the future growth of the technology, towards a mass deployment of secure distributed computing.

\section*{Acknowledgment}

The work of C.~Cicconetti was funded by the Horizon Europe Research and Innovation Programme action QUID (Grant Agreement no. 101091408) and project SERICS (PE00000014) under the MUR National Recovery and Resilience Plan funded by the European Union--NextGenerationEU.
The work of P.~Noviello and G.~.D.~Paduanelli was funded by EU, \textit{PON Ricerca e Innovazione} 2014--2020 FESR/FSC Project ARS01\_00734 QUANCOM.


\input{acronyms}

%
%






\end{document}

%% file: packages.tex
\usepackage{amsmath,amssymb,amsfonts}
\usepackage{algorithmic}
\usepackage{graphicx}
\usepackage{textcomp}
\usepackage{xcolor}
\usepackage[ruled,linesnumbered]{algorithm2e} 
\usepackage[nolist]{acronym}
\usepackage{soul}
\usepackage{marginnote}
\usepackage{url}
\usepackage[inline]{enumitem}
\usepackage{tabularx}
\usepackage{amsthm}
\usepackage{tikz}
\usepackage{tikz}
\usepackage{hyperref}
\hypersetup{
    colorlinks=true,
    linkcolor=blue,
    filecolor=magenta,      
    urlcolor=cyan,
    pdftitle={Overleaf Example},
    pdfpagemode=FullScreen,
    }
\usepackage{cleveref}

%% file: acronyms.tex
\begin{acronym}
  \acro{3GPP}{Third Generation Partnership Project}
  \acro{5G-PPP}{5G Public Private Partnership}
  \acro{AA}{Authentication and Authorization}
  \acro{ADF}{Azure Durable Function}
  \acro{AI}{Artificial Intelligence}
  \acro{API}{Application Programming Interface}
  \acro{AP}{Access Point}
  \acro{AR}{Augmented Reality}
  \acro{BGP}{Border Gateway Protocol}
  \acro{BSP}{Bulk Synchronous Parallel}
  \acro{BS}{Base Station}
  \acro{CDF}{Cumulative Distribution Function}
  \acro{CFS}{Customer Facing Service}
  \acro{CPU}{Central Processing Unit}
  \acro{DAG}{Directed Acyclic Graph}
  \acro{DHT}{Distributed Hash Table}
  \acro{DNS}{Domain Name System}
  \acro{ETSI}{European Telecommunications Standards Institute}
  \acro{FCFS}{First Come First Serve}
  \acro{FSM}{Finite State Machine}
  \acro{FaaS}{Function as a Service}
  \acro{GPU}{Graphics Processing Unit}
  \acro{HTML}{HyperText Markup Language}
  \acro{HTTP}{Hyper-Text Transfer Protocol}
  \acro{ICT}{Information and Communication Technologies}
  \acro{ICN}{Information-Centric Networking}
  \acro{IETF}{Internet Engineering Task Force}
  \acro{IIoT}{Industrial Internet of Things}
  \acro{ILP}{Integer Linear Programming}
  \acro{IPP}{Interrupted Poisson Process}
  \acro{IP}{Internet Protocol}
  \acro{ISG}{Industry Specification Group}
  \acro{ITS}{Intelligent Transportation System}
  \acro{ITU}{International Telecommunication Union}
  \acro{IT}{Information Technology}
  \acro{IaaS}{Infrastructure as a Service}
  \acro{IoT}{Internet of Things}
  \acro{JSON}{JavaScript Object Notation}
  \acro{K8s}{Kubernetes}
  \acro{KME}{Key Management Entity}
  \acro{KVS}{Key-Value Store}
  \acro{LCM}{Life Cycle Management}
  \acro{LCMP}{Life Cycle Management Proxy}
  \acro{LL}{Link Layer}
  \acro{LOCC}{Local Operations and Classical Communication}
  \acro{LTE}{Long Term Evolution}
  \acro{MAC}{Medium Access Layer}
  \acro{MBWA}{Mobile Broadband Wireless Access}
  \acro{MCC}{Mobile Cloud Computing}
  \acro{MEC}{Multi-access Edge Computing}
  \acro{MEH}{Mobile Edge Host}
  \acro{MEO}{MEC Orchestrator}
  \acro{MEPM}{Mobile Edge Platform Manager}
  \acro{MEP}{Mobile Edge Platform}
  \acro{ME}{Mobile Edge}
  \acro{ML}{Machine Learning}
  \acro{MNO}{Mobile Network Operator}
  \acro{NAT}{Network Address Translation}
  \acro{NISQ}{Noisy Intermediate-Scale Quantum}
  \acro{NFV}{Network Function Virtualization}
  \acro{NFaaS}{Named Function as a Service}
  \acro{NN}{Neural Network}
  \acro{OSPF}{Open Shortest Path First}
  \acro{OSS}{Operations Support System}
  \acro{OS}{Operating System}
  \acro{OWC}{OpenWhisk Controller}
  \acro{PMF}{Probability Mass Function}
  \acro{PPP}{Poisson Point Process}
  \acro{PU}{Processing Unit}
  \acro{PaaS}{Platform as a Service}
  \acro{PoA}{Point of Attachment}
  \acro{PPP}{Poisson Point Process}
  \acro{PQC}{Post Quantum Cryptography}
  \acro{QC}{Quantum Computing}
  \acro{QKD}{Quantum Key Distribution}
  \acro{QKDE}{QKD Entity}
  \acro{QoE}{Quality of Experience}
  \acro{QoS}{Quality of Service}
  \acro{RPC}{Remote Procedure Call}
  \acro{RR}{Round Robin}
  \acro{RSU}{Road Side Unit}
  \acro{SAE}{Secure Application Entity}
  \acro{SBC}{Single-Board Computer}
  \acro{SDK}{Software Development Kit}
  \acro{SDN}{Software Defined Networking}
  \acro{SJF}{Shortest Job First}
  \acro{SLA}{Service Level Agreement}
  \acro{SMP}{Symmetric Multiprocessing}
  \acro{SoC}{System on Chip}
  \acro{SLA}{Service Level Agreement}
  \acro{SRPT}{Shortest Remaining Processing Time}
  \acro{SPT}{Shortest Processing Time}
  \acro{STL}{Standard Template Library}
  \acro{SaaS}{Software as a Service}
  \acro{TCP}{Transmission Control Protocol}
  \acro{TLS}{Transport Layer Security}
  \acro{TSN}{Time-Sensitive Networking}
  \acro{TN}{Trusted Node}
  \acro{UDP}{User Datagram Protocol}
  \acro{UE}{User Equipment}
  \acro{URI}{Uniform Resource Identifier}
  \acro{URL}{Uniform Resource Locator}
  \acro{UT}{User Terminal}
  \acro{VANET}{Vehicular Ad-hoc Network}
  \acro{VIM}{Virtual Infrastructure Manager}
  \acro{VR}{Virtual Reality}
  \acro{VM}{Virtual Machine}
  \acro{VNF}{Virtual Network Function}
  \acro{WLAN}{Wireless Local Area Network}
  \acro{WMN}{Wireless Mesh Network}
  \acro{WRR}{Weighted Round Robin}
  \acro{YAML}{YAML Ain't Markup Language}
\end{acronym}